High-pressure study of superconducting and non-superconducting single crystals of the same nominal composition $Rb_{0.8}Fe_2Se_2$

M. Gooch,[1] B. Lv,[1] L. Z. Deng,[1,2] T. Muramatsu,[1] J. Meen,[1,3] Y. Y. Xue,[1,2] B. Lorenz[1,2] and C. W. Chu[1,2,4]

[1]Texas Center for Superconductivity, University of Houston, Houston TX  77204-5002, USA

[2]Department of Physics, University of Houston, Houston TX  77204-5005, USA

[3]Department of Chemistry, University of Houston, Houston TX  77204-5003, USA

[4]Lawrence Berkeley National Laboratory, 1 Cyclotron Road, Berkeley CA  94720, USA

Abstract

Two single crystalline samples with the same nominal composition of $Rb_{0.8}Fe_2Se_2$ prepared via slightly different precursor routes under the same thermal processing conditions were investigated at ambient and high pressures. One sample was found superconducting with a $T_c$ of ~31 K without the previously reported resistivity-hump and the other was unexpectedly found to be a narrow-gap semiconductor. While the high pressure data can be understood in terms of pressure-induced variation in doping, the detailed doping effect on superconductivity is yet to be determined.

I. Introduction

The recent discovery of the Fe-based chalcogenide 30 K superconductors $A_xFe_2Se_2$, where A = alkaline element K, Rb, Cs or Tl, that possess the tetrahedrally coordinated corner sharing FeSe-layers have generated great interest.[1-4] Because of the similarity of the crystal structures and the close chemical proximity between these chalcogenides and their superconducting pnictide counterparts,[5] many experiments have since been carried out to explore the occurrence of superconductivity in Fe-chalcogenides. These efforts give hope for shedding light on the superconducting mechanism in Fe-chalcogenides and -pnictides in particular, and in high temperature superconducting cuprates in general.

In contrast to the pnictides, the simplest Fe-based superconducting chalcogenide, *i.e.* $FeSe_x$ with x < 1, was found to crystallize in the simple binary layer anti-PbO structure (P4/nmm), which displays a superconducting transition temperature ($T_c$) of 8 K.[6] The $T_c$ of $FeSe_x$ has been raised either to ~ 15 K[7] by partial replacement of Se by Te or to ~ 37 K[8] by pressurization at 4 GPa. Until the recent advent of $A_xFe_2Se_2$, no effort has been successful for stabilizing the chalcogenides in a phase with a more complex layer-structure similar to the layer-pnictides, e.g. the 1111 phase[1,2] [RFeAsO, where R = rare earth] with the ZrCuSiAs structure (P4/nmm) and a maximum $T_c$ of 57 K when doped ; the 122 phase[9,10] [AeFe$_2$As$_2$ or AFe$_2$As$_2$, where Ae = alkaline earth or A= alkaline] with the ThCr$_2$Si$_2$ structure (I4/mmm) and a maximum $T_c$ of 38 K when doped; or the 111 phase[11,12,13] [AFeAs, where A = alkaline] with the PbFCl structure (P4/nmm) and a maximum $T_c$ of 25 K when doped. A $T_c$ between 29 and 33 K has recently been discovered in A$_x$Fe$_{2-y}$Se$_{2+z}$, when 0.6 < x < 0.9, 0 < y < 0.7, and 0 < z < 0.05, almost independent of A = K,[1,2] Rb,[4] or Cs.[3] These superconducting chalcogenides exhibit a similar $T_c$ and crystal structure to those of the 122 phase of the Fe-pnictides, strongly suggesting a similar major role of Fe in the occurrence of superconductivity in the chalcogenides and pnictides. However, band structure calculations[14,15] suggest that they are more like the binary-layer anti-Pb-structured FeSe$_x$ intercalated with A between the FeSe layer and that the superconductivity evolves from the antiferromagnetic semiconductor parent compound; this suggestion is further supported by experimental[2,16] results. This is in stark contrast to the pnictides, where the superconducting phase evolves from the antiferromagnetic semimetal parent compound.[5,17] The absence of anomalies in thermoelectric power is proposed[18] to indicate that very little Fermi surface nesting takes place in these compounds, seemingly consistent with the absence of a hole pocket in the zone center, as concluded from the band calculations[19] and from ARPES,[19,20] and thus ruling out Fermi surface nesting as the major cause for superconductivity.

Experiments on this newly discovered family of superconducting Fe-chalcogenides show the sensitive dependence of superconductivity on the stoichiometry and synthesis conditions.[21,22] Detailed dependences are yet to be revealed. Recognizing the effects of pressure on the superconducting properties of compounds without introducing the complexity of chemical substitution, several high pressure studies[23-27] have been carried out on the superconducting A$_x$Fe$_{2-y}$Se$_{2+z}$ single crystals for A = K[23,24,25,27] and Cs[25-27]. As for A = Rb, only pressure dependent synchrotron power diffraction measurements have been reported.[27] Here we report the resistive and magnetic results of both the superconducting and non-superconducting single crystalline samples of the same nominal Rb$_{0.8}$Fe$_2$Se$_2$, for ambient and high pressure. The superconducting and non-superconducting single crystalline samples were both prepared by the self flux technique via similar thermal steps but slightly different precursor routes. Under pressure, the $T_c$ of the superconducting sample increases slightly and then continually decreases above 1.75 GPa, until superconductivity is no longer detected at ~5.6 GPa. The non-superconducting sample behaves like a semiconductor and exhibits a magnetic transition at ~ 150 K. However, no sign of superconductivity was detected up to 10 GPa in the non-superconducting sample. The results will be presented and compared with those of the other members of the family.

II. Experimental

We have determined the resistance of the samples with a four-lead technique using the LR-700 AC Resistance Bridge at pressures up to ~ 1.8 GPa in a BeCu clamp cell and with the pseudo-four lead method deploying a home-made diamond anvil cell (DAC) under pressure up to 10 GPa. The superconducting transition temperature, $T_c$, was determined by the midpoint or 50% drop seen in the resistance from the onset of superconductivity. The dc magnetic susceptibility was measured at ambient pressure using the Quantum Design SQUID Magnetometer in a field up to 5 T down to 2 K. The temperature was determined by various thermometers depending on the temperature range and the different probes used with a general uncertainty ± (1-10) mK. The pressure was determined by the superconducting Pb-scale at low temperature up to 1.8 GPa with a resolution of ± 0.02 GPa and by the ruby fluorescence scale at room temperature up to 10 GPa with a resolution of ± 0.2 GPa. The pressure medium used in the BeCu clamp cell is a mixture of Fluorinert 70 and 77 while that in the diamond anvil cell is NaCl.

III. Results and Discussions

Single crystalline samples of $Rb_{1-x}Fe_{2-y}Se_2$ are grown from their own flux following different precursor routes but similar thermal steps. Samples of sizes as large as ~ 10 x 10 x 0.2 $mm^3$ were obtained. Details of the synthesis process will be published elsewhere.[28] It is interesting to note that samples with the same nominal composition and processed under similar conditions often end up with completely different electrical properties, although displaying the same crystal structure. In this investigation, we shall focus only on two such kinds of samples with the same nominal composition of Rb/Fe/Se = 0.8/2/2 synthesized under the same conditions but via different precursor routes: one is superconducting (A) and the other is semiconducting (B). Sample A was prepared by reacting Rb (99.75% pure from Alfa) with the appropriate amount of our preformed FeSe (99.99+% pure Fe from Aldrich and 99.999% pure Se from Alfa) while Sample B by direct reaction of Rb, Fe, and Se. Stoichiometric amounts of starting materials were placed in an alumina crucible inside a silica tube, which was sealed under reduced Ar atmosphere. The tube was then sealed inside another larger silica tube under vacuum. The assembly was finally put inside the box furnace, heated up to 1020˚C in 8 hours, and then very slowly cooled down to 750˚C at a rate of 6˚C/hour. The furnace was then turned off and the samples were cooled naturally to room temperature. The crystal structure was determined by a Siemens SMART CCD single-crystal diffractometer and a Panalytical X'pert Diffractometer. No impurity phases could be detected within the resolution of the X-ray spectra for powders as well as single crystals, limiting the possible content of other phases to a few percent. The chemical composition was measured by a Wavelength Dispersive Spectrometer (WDS). All single-crystal diffraction spots over $0 < 2\theta \leq 60^o$ with a relative strength > 1% can be indexed with a $ThCr_2Si_2$ structure except for a set of weak peaks on the left-side of the *00ℓ* lines. The extracted lattice parameters are [a = b = 8.758(1) Å and c = 14.570(5) Å] and [a = b = 8.748(2) Å and c = 14.602 (6) Å] for samples A and B, respectively. The θ-2θ XRD patterns, however, show weak reproducible peaks on the left side of the *008, 0010* and *0012* lines at large 2θ (marked by stars in Fig. 1) and the separations between them increase with *ℓ systematically*, suggesting some yet-to-be identified intergrowth-like microstructures with similar a and b, but different c. These peaks, however, bear no apparent correlations with superconductivity and are relatively weak. The WDS analysis shows the chemical compositions are Rb/Fe/Se = 0.93(2)/1.70(2)/2.00 and 0.90(1)/1.78(1)/2.00 for samples A and B, respectively. The uncertainties cited are the statistical fluctuation over six well-separated points. The possible systematic errors are eliminated through calibration procedures. The estimated precision is better than 0.5%.

As shown above, in spite of the same nominal compositions and the same thermal procedure of sample synthesis, slight variations in lattice parameters and chemical compositions were detected, presumably due to the different precursor routes adopted for samples A and B. However, more surprising is the drastically different electric and magnetic properties of the two samples observed. Sample A is metallic and becomes superconducting at low temperature as evident in its temperature dependence of resistivity ρ(T), shown in Fig. 2a. On cooling, ρ decreases continuously, exhibits a large negative curvature between 300 K and 125 K and drops precipitously to a superconducting state with an onset $T_c$ of 31 K (see inset, Fig. 2a) but without the ρ-hump previously reported in $A_xFe_{2-y}Se_2$ between ~ 120 and 300 K.[1-4] Although ordering of the Fe-vacancies has been suggested,[29] the origin of the hump remains unclear.

On the other hand, sample B behaves as a narrow-gap semiconductor as predicted, as shown in Fig. 2b. Its ρ(T) has a room temperature value of 0.428 Ω-cm, about 3 times smaller than that of Sample A, and increases exponentially as temperature decreases with an activation energy of 5 meV below ~ 160 K. The 10 Oe magnetic susceptibilities χ(T) of samples A and B are displayed in Figs. 3a and 3b, respectively. For sample A at 10 Oe, the zero-field-cooled $χ_{ZFC}$(T) shows a small positive background (~ 6·10$^{-3}$ emu/cm$^3$) and it turns negative at ~ 31 K, signaling the onset of the superconducting transition, in agreement with the ρ(T) in Fig. 2a. The diamagnetic signal, corrected for the demagnetization effects, $4πχ_{ZFC}$ ≈ -0.55 at 5 K indicates bulk superconductivity in our sample A. However, χ(T) for sample B in Fig. 3b indicates the absence of a superconducting transition. It is nearly constant at higher temperatures ( ≈ .111 emu/cm$^3$) and it displays a sharp increase at about 130 K (Fig. 3b). The M(H) loops at 2 K, 100 K, and 300 K are shown in the inset of Fig. 3b. The strongly nonlinear increase of M(H) and a small field hysteresis at low fields (coercive field < 360 Oe) suggest the existence of a weak ferromagnetic moment at all temperatures below 300 K. The magnetic anomaly of χ(T, 10 Oe) at 130K in sample B may indicate another change in the magnetic system, possibly associated with the ordering of recently proposed magnetic clusters surrounding the Fe-vacancies.[15] However, the presence of a small amount of $Fe_7Se_8$ impurity phase, below the detection limit of the X-ray spectra, cannot be ruled out. $Fe_7Se_8$ has a ferromagnetic moment below room temperature and it exhibits a spin reorientation close to 130 K.[30,31] The large magnitude of this moment could make a significant contribution to the moment of sample B, shown in Fig. 3b, if only a few percent of $Fe_7Se_8$ are present.

The apparent differences in the physical properties of samples A (superconducting) and B (semiconducting) have to be explained by subtle differences of their electronic structure and charge densities at the iron site. With the compositions derived from WDS measurements, the iron valences of both samples indeed differ by about 0.06 (assuming $Rb^{1+}$ and $Se^{2-}$). Small differences in the electronic structure should also be reflected in the thermoelectric power since this quantity is sensitive to the charge density and the Fermi surface topology. The thermoelectric power results S(T) for the two samples are shown in Fig. 4. They are positive for both samples at room temperature, change sign at around 200 K, and exhibit a broad minimum at around 100 K. However, on further cooling, S(T) for sample A becomes zero, indicative of the entering into a superconducting state at ~ 31 K, but not for sample B, consistent with the ρ(T) and χ(T) data. The small positive values of S(T) at 300 K and the sign change at 200 K indicate that the Fermi surface is more complex and may involve hole as well as electron like carriers competing for the sign of S at different temperatures. A similar sign change of S(T) was reported for a semiconducting sample of the sister compound $K_xFe_{2-y}Se_2$.[32] Nevertheless, the differences in the slopes of S(T) for samples A and B confirm that both samples are electronically different, consistent with the resistivity and magnetization data. In particular, the diffusion (linear) part of S(T) is directly related to the Fermi surface topology through the energy-dependent electronic conductivity.[33] The small change of the doping state (only 0.06 electrons per Fe) and the dramatic differences in the physical properties indicate that samples A and B are close to but on different sides of a metal-insulator phase boundary. This conjecture is supported by a recent study of the system $Rb_xFe_{2-y}Se_2$ for a variety of x and y values, which has suggested that the superconducting phase is stable over a narrow doping range and is sandwiched between an insulating and a semiconducting phase.[34]

The observation of completely different electric and magnetic properties displayed by the single crystalline samples of $Rb_xFe_{2-y}Se_2$ prepared with the same nominal compositions (Rb/Fe/Se = 0.8/2/2) and by the same method with the same thermal steps is extremely intriguing. Although the XRD and WDS-data show slight differences in lattice parameters (~ 0.1% in a and b; ~ 0.22% in c) and compositions [Rb/Fe/Se = 0.93(2)/1.70(2)/2.00(1) and 0.90(1)/1.78(1)/2.00(1)], the chemical composition spread appears to fall within a narrow range close to reported data for $Rb_xFe_{2-y}Se_2$.[4,29,34] It should be noted that most reported their nominal compositions but few did composition analysis. The results suggest that superconductivity in this family of Fe-chalcogenides depends sensitively not just on doping as reflected in the chemical composition but also on the defects (Fe and/or Rb vacancies) and their state (ordered and/or disordered) present in the samples.[34] With respect to doping, they are very different from the Fe-pnictide family where a rather large doping range exists, i.e. $Sr_{1-x}K_xFe_2As_2$ for $0 \leq x \leq 1$.[35] Our preliminary study indicates that the doping range for superconductivity appears to be extremely narrow.

The pressure effect on the R(T) of sample A is shown in Fig. 5. It is evident that, in general, pressure suppresses R(T) progressively, which is in agreement with other published reports on sister compounds[23-27]; however sample A lacks the hump anomaly in the resistance that has been reported by others. The inset of Fig. 5 shows the pressure effect on R(T) at low temperature. The $T_c$ increases only slightly from ~29.3 K at ambient to ~30.1 K at ~1 GPa and decreases above, as seen in Fig. 6 (inset). For pressures above ~1.75 GPa, a DAC was used. At ~5.6 GPa, superconductivity is no longer seen. The suppression of superconductivity under pressure was also reported previously at ~8.7 GPa[23] in $K_{0.8}Fe_{1.7}Se_2$ and at ~ 8 GPa[25] in $Cs_{0.8}Fe_2Se_2$. It should be noted that our sample has a lower critical pressure and no hump in the resistance, though the overall trend is similar. The overall results can be understood in terms of pressure-induced doping as has been demonstrated previously in cuprate high temperature superconductors.[36]

For the non-superconducting sample B, the overall ρ(T) is drastically suppressed by pressure. No superconductivity was detected in sample B up to 10 GPa, in contrast to the fact that almost all non-superconducting Fe-pnictides become superconducting under pressures. However, it should be noted that the possibility of the contact resistance associated with the pseudo-four lead technique employed to mask the superconducting transition cannot be completely ruled out at the present time.

Based on the recent observation made on $K_{0.8}Fe_{1.7}Se_2$, where pressure suppresses the amplitude of the ρ-hump and superconductivity and the ρ-hump disappear simultaneously at ~ 9.5 GPa, a possible correlation between the ρ-hump with superconductivity has also been conjectured.[23] The absence of the ρ-hump up to 300 K in our $Rb_{0.93}Fe_{1.70}Se_{2.00}$ and its $T_c$ of ~ 30 K, similar to that of $A_xFe_{2-y}Se_{2-z}$, led us to the suggestion that the ρ-hump is not related to the superconductivity in this compound family and that the simultaneous pressure-induced complete suppression of both $T_c$ and the ρ-hump detected in $K_{0.8}Fe_{1.7}Se_2$ may just be coincidental. The large variation of the temperature of the ρ-hump with a more or less constant $T_c$ provides further support to the above suggestion.

IV.  Conclusion

Two single crystalline samples of the same nominal composition $Rb_{0.8}Fe_2Se_2$ were prepared using the self-flux technique via two different precursor routes following the same thermal history. Both samples display the same $ThCr_2Si_2$-structure with only slight differences in lattice parameters and the actual chemical composition as revealed by the WDS analysis. Although the difference in the final chemical composition falls within a narrow range, one was superconducting with a $T_c$ ~ 30 K, while the other behaves like a narrow gap semiconductor with activation energies of 5 meV below 160 K. In spite of the similar $T_c$, our superconducting sample does not show the ρ-hump between $T_c$ and 300 K, suggesting that superconductivity in Fe-chalcogenides is not related to the ρ-hump in contrast to previous contention. A magnetic transition is detected at 130 K in the non-superconducting sample.  The origin of this anomaly could be attributed to a possible magnetic ordering of the Fe-vacancies; however, it cannot be excluded that this anomaly could also be related to impurity phase(s) that are below the detection limit of powder and single crystal X-ray spectra. Upon increasing pressure, $T_c$ of the superconducting sample was found to increase slightly initially followed by a decrease above 1 GPa, and superconductivity disappears above ~5.6 GPa, similar to other members of the family. This can be understood in terms of a pressure-induced change in doping as for cuprate high temperature superconductors.

The superconducting properties in $A_xFe_{2-y}Se_2$ are closely affected by the chemical composition, the doping state, and the defects of the sample. Those parameters depend sensitively on the detailed sample processing conditions, although the superconducting $T_c$s are almost independent of the alkali metal. A systematic study on the influence of chemical composition and defects on the superconducting and magnetic properties of $A_xFe_{2-y}Se_2$ is warranted to help unravel the mystery of superconductivity in Fe-chalcogenides.

Acknowledgement

The work in Houston is supported in part by US Air Force Office of Scientific Research contract FA9550-09-1-0656, Department of Energy subcontract 4000086706 through ORNL, AFRL subcontract R15901 (CONTACT) through Rice University, the T. L. L. Temple Foundation and the John J. and Rebecca Moores Endowment, the State of Texas through TCSUH. The work at Lawrence Berkeley National Laboratory is supported by the Director, Office of Science, OBES, DMSE, and Department of Energy.

Figure Captions

Fig. 1: The θ-2θ XRD patterns of samples A and B. The stars indicate unidentified, possibly intergrowth microstructures with similar a and b but different c-parameters.

Fig. 2: a) The $\rho(T)$ of sample A with the inset showing the low temperature $\rho(T)$. b) The $\rho(T)$ of sample B with the inset showing the $\ln[\rho(T)]$ vs. $1/T$ behavior.

Fig. 3: a) The 10 Oe magnetic susceptibility after demagnetization correction of sample A. The solid symbols are zero-field-cooled data and the open symbols, the field-cooled ones. b) The magnetic moment of sample B at 10 Oe. Inset: M-H loops of sample B at 2 K, 100 K, and 300 K (from top to bottom).

Fig. 4: The thermoelectric power of samples A and B.

Fig. 5: The resistance of sample A at different pressures. Inset shows the low temperature R(T). The numbers in both parts represent the sequence of the experimental runs: 1—0; 2—0.31 GPa; 3—0.91 GPa; 4—1.75 GPa; 5—0.64 GPa; 6—1.4 GPa; 7—1.47 GPa.

Fig. 6: The pressure dependence for sample A, $T_c(P)$, where the solid symbols, 1-7, represent the BeCu clamp cell data and the open symbols, a-e, represent the DAC data. Inset: the pressure dependence at lower pressures obtained with a BeCu cell. The numbers indicate the order of the applied pressure.

Fig. 1

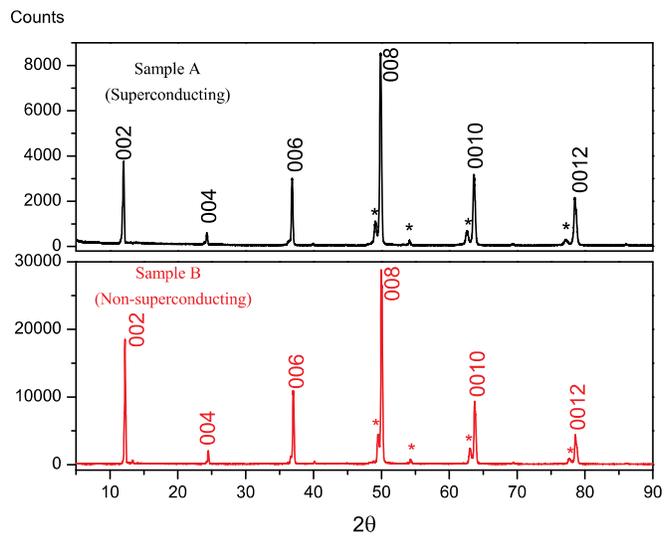

Fig. 2

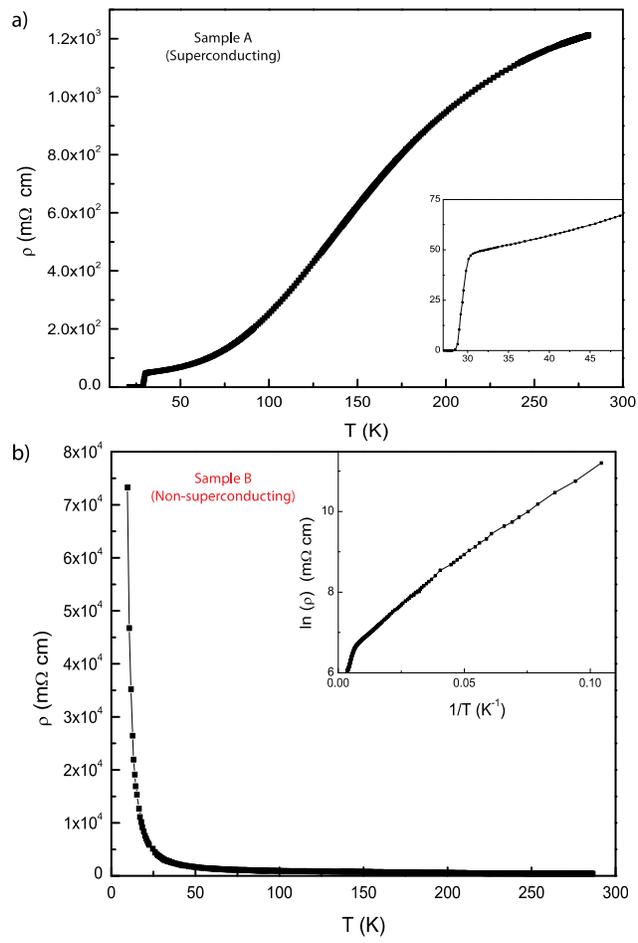

Fig. 3

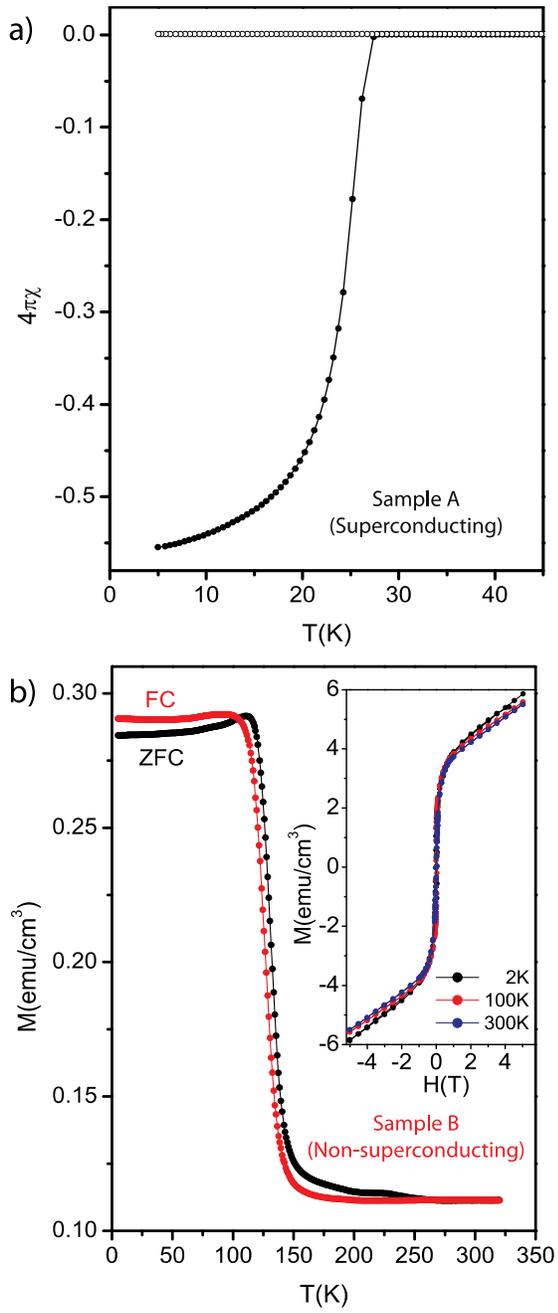

Fig. 4

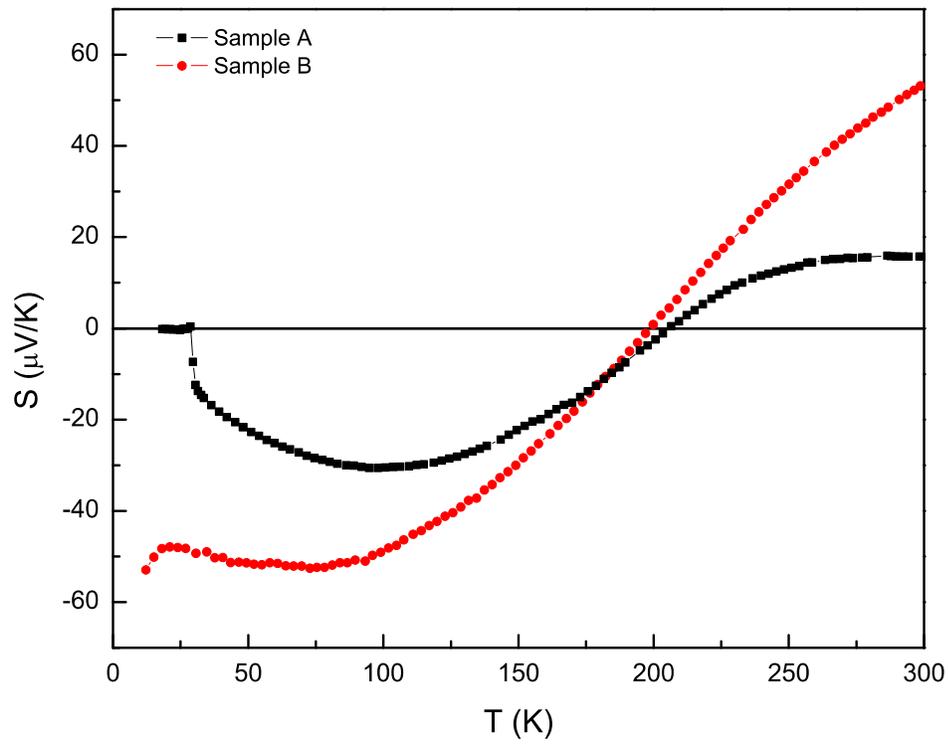

Fig. 5

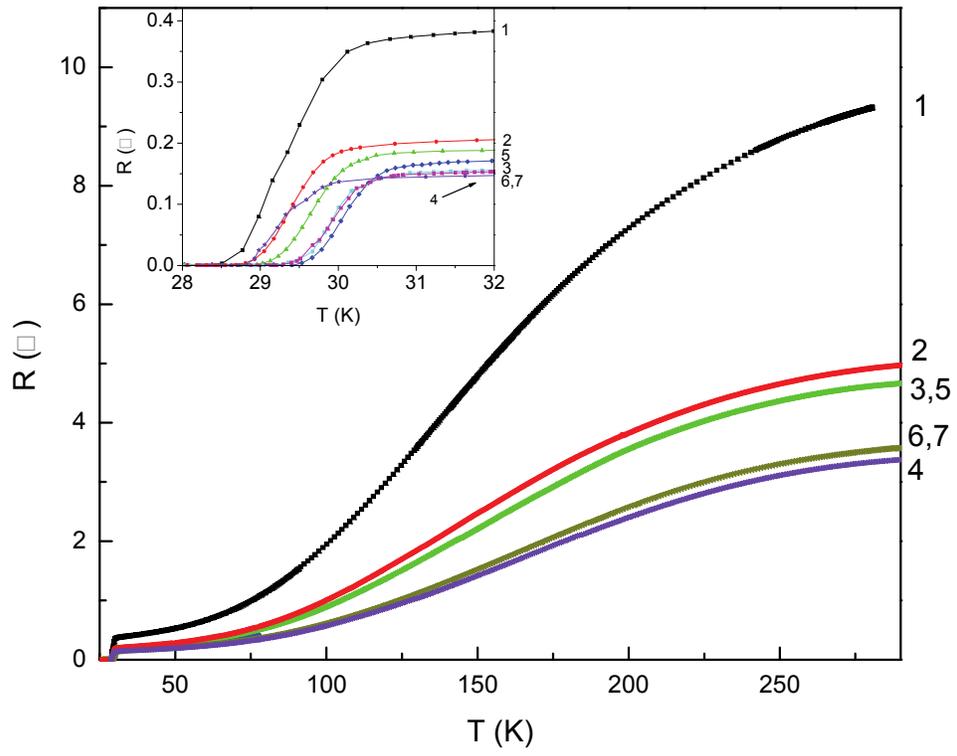

Fig. 6

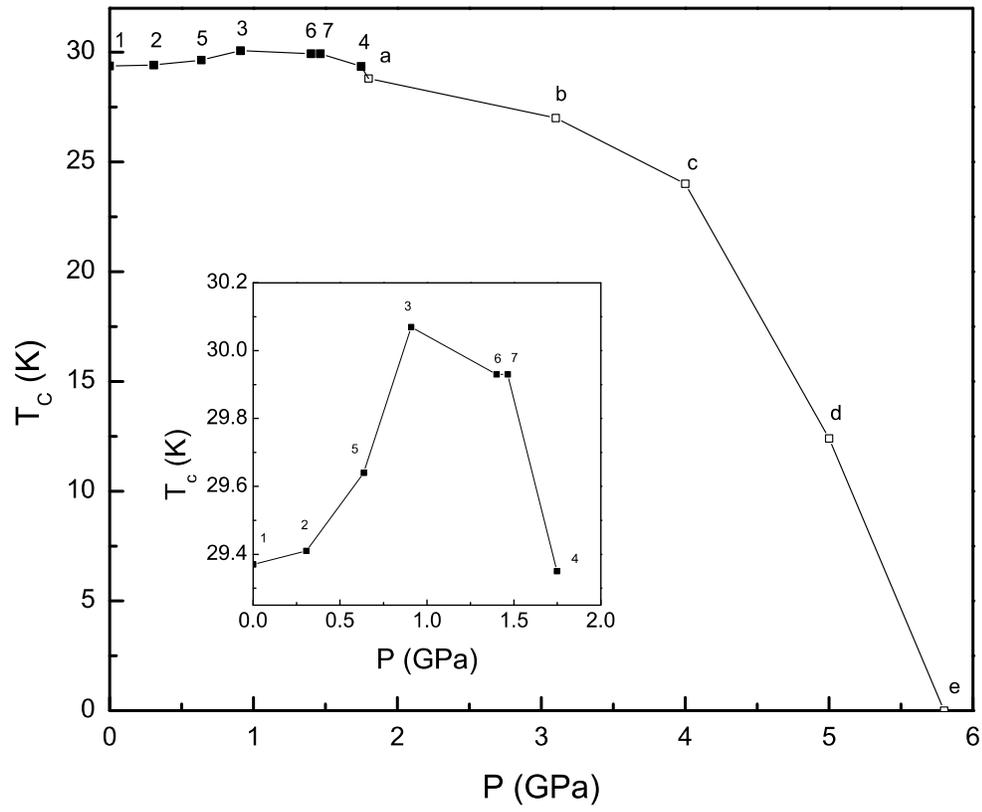

15